\documentclass[a4paper, 11pt]{article}
\usepackage[top=18truemm,bottom=25truemm,left=25truemm,right=25truemm]{geometry} 

\usepackage{indentfirst} 

\makeatletter
\def\section{\@startsection {section}{1}{\z@}{-3.5ex plus -1ex minus -.2ex}{2.3 ex plus .2ex}  {\large\bf}}
\makeatother
\usepackage{secdot} 

\makeatletter
\def\subsection{\@startsection {subsection}{1}{\z@}{-3.5ex plus -1ex minus -.2ex}{2.3 ex plus .2ex}{\small\bf}}
\makeatother

\usepackage{amsmath, amssymb} 

\usepackage[dvipdfmx]{graphicx}
\usepackage{float} 
\graphicspath{{./}} 
\usepackage[labelsep=period, figurename=Fig. ,tablename=Table]{caption} 
\usepackage{subcaption} 
\usepackage{booktabs}
\usepackage{multirow}
\usepackage{tabularx}
\newcolumntype{C}{>{\centering\arraybackslash}X}

\usepackage{cite}
\usepackage{url}

\usepackage[T1]{fontenc} 
\usepackage{ulem}
\usepackage{comment} 
\usepackage{color}
\usepackage{ascmac}
\usepackage{fancybox}
\usepackage{seqsplit} 

\title{Validation of a motion model for soccer players' sprint\\ by means of tracking data}

\usepackage{authblk}
\author[1]{Takuma Narizuka\thanks{{\it E-mail address}: pararel@gmail.com (T. Narizuka).\\
}}
\author[2]{Kenta Takizawa}
\author[3]{Yoshihiro Yamazaki}
\affil[1]{Faculty of Data Science, Rissho University, Kumagaya, Saitama 360-0194, Japan}
\affil[2]{Department of Physics, Faculty of Science and Engineering, Chuo University, Bunkyo, Tokyo 112-8551, Japan}
\affil[3]{Department of Physics, School of Advanced Science and Engineering, Waseda University, Shinjuku, Tokyo 169-8555, Japan}
\date{}

\begin{document}
	\maketitle
	\baselineskip 14pt

\begin{abstract}
\baselineskip 14pt
In soccer game analysis, the widespread availability of play-by-play and tracking data has made it possible to test mathematical models that have been discussed mainly theoretically.
One of the essential models in soccer game analysis is a motion model that predicts the arrival point of a player in $ t $ s.
Although many space evaluation and pass prediction methods rely on motion models, the validity of each has not been fully clarified.
This study focuses on the motion model proposed by Fujimura and Sugihara (Fujimura-Sugihara model) under sprint conditions based on the equation of motion.
A previous study indicated that the Fujimura-Sugihara model is ineffective for soccer games because it generates a circular arrival region.
This study aims to examine the validity of the Fujimura-Sugihara model using soccer tracking data.
Specifically, we quantitatively compare the arrival regions of players between the model and real data.
We show that the boundary of the player's arrival region is circular rather than elliptical, which is consistent with the model.
We also show that the initial speed dependence of the arrival region satisfies the solution of the model.
Furthermore, we propose a method for estimating valid kinetic parameters in the model directly from tracking data and discuss the limitations of the model for soccer games based on the estimated parameters.
\end{abstract}

\section{Introduction}
Over the past decade, various types of data have become available for sports data analysis.
In particular, the prevalence of play-by-play \cite{Pappalardo2019} and tracking data \cite{Pettersen2014} for soccer games has enabled new analyses that were previously impossible \cite{Gudmundsson2017}.
Some typical examples include ball-passing network analysis \cite{Duch2010, Buldu2019}, formation analysis \cite{Bialkowski2014, Narizuka2019}, and space evaluation \cite{Fernandez2018, Spearman2018, Narizuka2021}.
Such research topics require a variety of statistical analysis methods, including network theory, computational geometry, and machine learning.
In particular, machine learning is an effective tool for soccer game analysis because soccer produces very complex behaviors from simple unified rules and has a huge accumulation of data.
Soccer games are a good laboratory to develop cutting-edge machine learning techniques \cite{Google2020, Tuyls2021}.
Statistical properties of player interactions and collective motions are also hot topics in soccer game analysis.
In terms of statistical physics, soccer players can be regarded as self-propelled particles \cite{Sumpter2016}.
We can apply some techniques developed for characterizing the self-propelled particles, including flocks of birds or fish schools \cite{Vicsek2012}.
The examples include the detection of highly correlated segments using directional correlation functions \cite{Marcelino2020}, the characterization of order-disorder transition \cite{Narizuka2017}, and the modeling of soccer players' motion by the self-propelled player model \cite{Alguacil2020}.
Furthermore, the dynamics of player interactions and ball passing in soccer games are described as stochastic processes, such as the Markov chain \cite{Narizuka2014} and the first-passage process \cite{Chacoma2020}.
The widespread use of these new data has also made it possible to test mathematical models that have mainly been discussed theoretically in soccer game analysis.
One of the essential models in soccer games is a motion model, which calculates the arrival point of a player in $ t $ s based on the current location and velocity.
The two practical applications of motion models in soccer games are space evaluation and pass prediction.
In space evaluation, a fundamental concept is the ``dominant region,'' defined as the region where a specific player can arrive before other players \cite{Taki2000, Gudmundsson2017}.
In general, we can estimate the dominant region of each player by comparing the arrival times of all players to each location in the field.
In addition, the outcome of a given pass can be estimated by calculating the arrival time of each player on the ball trajectory \cite{Fujimura2005, Spearman2017, Alguacil2020, Anzer2022}.
Because the motion model can calculate the arrival times of a player to a specific location, it is essential for soccer game analysis.
Thus far, various motion models have been proposed.
The simplest motion model assumes uniform linear motion for all players, resulting in the Voronoi region as the dominant region.
More realistic models have also been proposed; for example, Taki and Hasegawa assumed uniform accelerated motion of players \cite{Taki1996, Taki2000}, and Fujimura and Sugihara considered viscous resistance \cite{Fujimura2005}.
These motion models are based on the equation of motion and are often referred to as the ``physics-based motion model.''
They provide the arrival points of players under specific conditions such that each player moves to all locations by sprinting.
Another type of motion model calculates the arrival points of players based on machine learning \cite{Gudmundsson2014, Brefeld2019, Caetano2021}.
This ``probabilistic motion model'' can predict realistic arrival points based on previous tracking data, though it is costly for learning.
We note that because the tracking data used for learning include various running patterns other than a sprint, the predicted arrival point does not necessarily mean the position arrived by sprinting.
Both physics-based motion models and probabilistic motion models aim to predict the arrival point of players.
Physics-based motion models also play a role in elucidating the principles of players' movement laws.
This study focuses on the motion model proposed by Fujimura and Sugihara (hereinafter Fujimura-Sugihara model) in sprint conditions to elucidate soccer players' movement laws during sprinting.
Specifically, we aim to investigate the validity and limitations of the Fujimura-Sugihara model based on soccer tracking data.
We stress that the motion model in the sprint condition is significant in situations where we calculate the minimum arrival time of players to each location.
The minimum arrival time is utilized to evaluate which player can receive the ball \cite{Fujimura2005, Spearman2017} and to quantify the degree of safety and sparsity of each location in the field \cite{Narizuka2021}.
Thus, the present study provides an essential basis for various applied analyses.
In our investigation, we first focused on the shape of the arrival region of players predicted by the motion model.
The Fujimura-Sugihara model generates a circular arrival region \cite{Fujimura2005}.
However, it has been pointed out that the arrival region should be elliptical, considering the difference in the acceleration ability of players depending on the direction of motion \cite{Brefeld2019, Fernandez2018}.
We show that the circular arrival region is obtained from soccer tracking data, and the arrival region's initial speed dependence also satisfies the Fujimura-Sugihara model's solution.
Next, we propose a method to estimate the kinetic parameters of the Fujimura-Sugihara model.
Contrary to previous experiment-based methods \cite{Fujimura2005}, this method can estimate valid parameters directly from tracking data.
Finally, we discuss the limitations of the Fujimura-Sugihara model for soccer games based on the time dependence of the estimated kinetic parameters.

\section{Method}

\subsection{Data}
The datasets used in this study were from 54 soccer games in the top division of the Japan Professional Football League (J1 League).
Each game was played by 18 teams in 2016.
The primary data of our dataset is absolute positional coordinates $ (x, y) $ of all players every 0.04 s, collected by using the TRACAB system \cite{tracab}.
The $ x $ and $ y $ coordinates are considered to contain an error of $ \pm 1 $ m by assessing the accuracy of the TRACAB system.
These datasets were provided by DataStadium Inc., Japan, which was authorized to collect and sell these data under a contract with the J League \cite{DataStadium}.
This contract also ensures that the use of relevant datasets does not infringe on the rights of players and clubs belonging to the J League.
Although the datasets were proprietary, we received explicit permission from DataStadium Inc. for use in this study.
This study's Data analyses and visualizations were performed using Python packages on an iMac Pro system with a 3-GHz 10-Core Intel Xeon W processor and 128 GB of memory.

\subsection{Preliminary Analysis}
\begin{figure}[htbp]
	\centering
	\includegraphics[width=14cm]{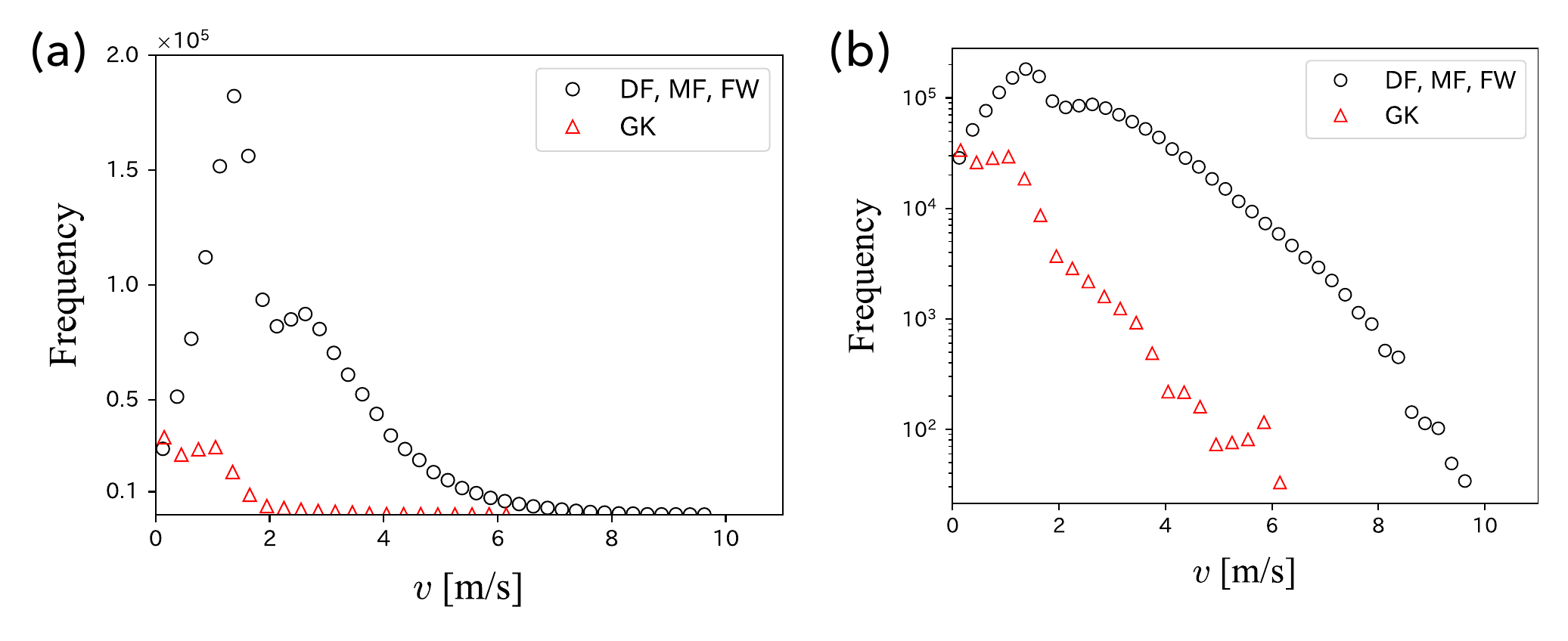}
	\caption{Speed distributions of soccer players. (a) Linear plot and (b) semi-logarithmic plot.}
	\label{dist_v}
\end{figure}

\begin{figure}[htbp]
	\centering
	\includegraphics[width=8cm]{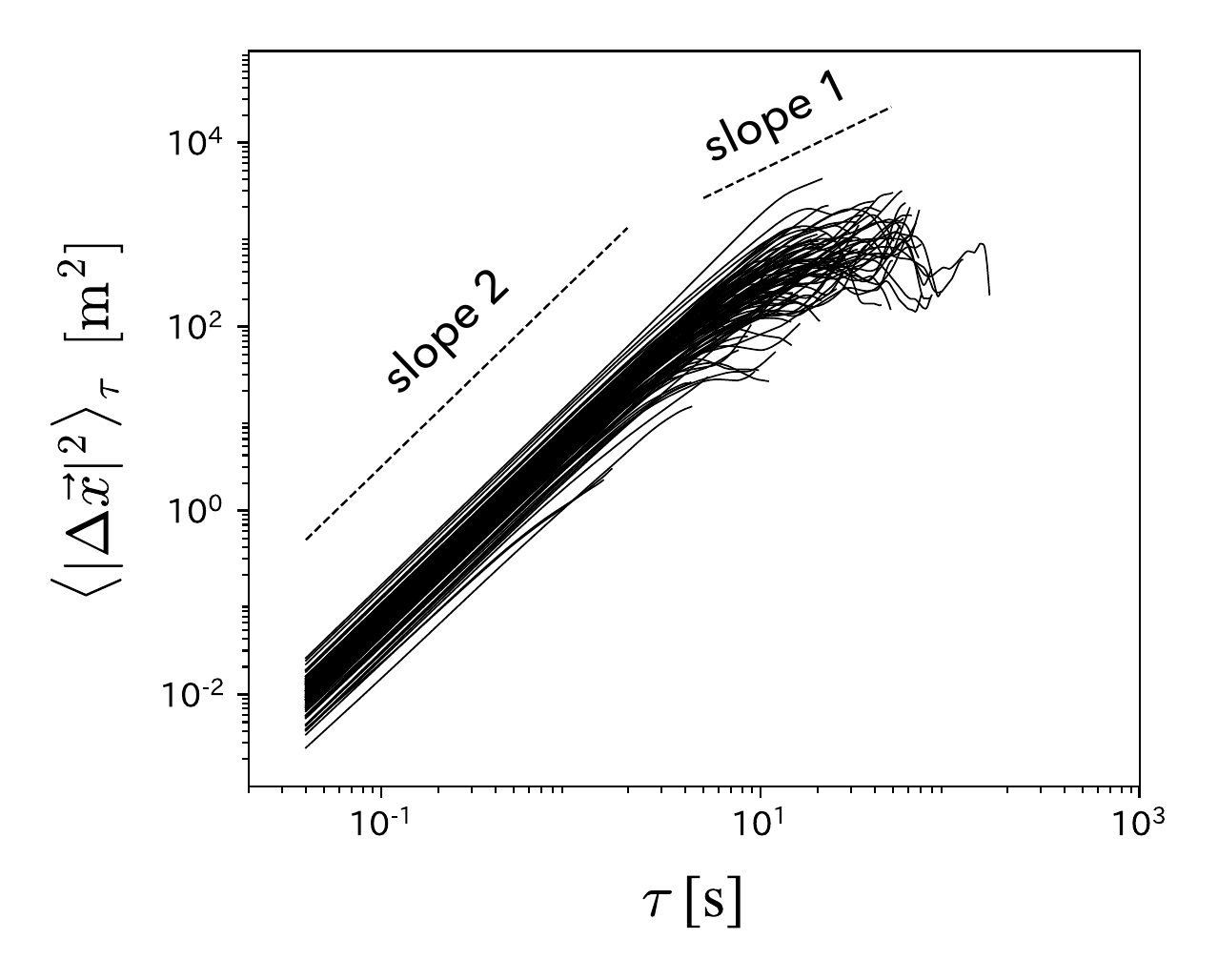}
	\caption{$ \tau $ dependence of the mean squared displacement of soccer players. The graph is shown in double logarithmic scale.}
	\label{msd}
\end{figure}

We summarize the basic properties of soccer players' motions.
The player's velocity $ \vec{v}(t) $ was calculated as the difference between the current position and the position 1 s ago.
Figure \ref{dist_v}(a) shows the speed distributions for the goalkeeper and the other players obtained from a game.
It is found that all players have a peak at $ v \simeq 1 $ m/s, whereas the non-goalkeepers have a second peak at $ v \simeq 3 $ m/s.
These two peaks correspond to walking and jogging, respectively.
Fig. \ref{dist_v}(b) shows that the speed distribution of the non-goalkeepers is more widely distributed than that of the goalkeeper, and both decay almost exponentially.
Because the sprint is observed mainly for players other than the goalkeeper, we exclude goalkeepers in the following analysis.
We also examine mean squared displacement (MSD) to characterize players' trajectories.
When a player at location $ \vec{x}(t) $ at $ t $ moves to location $ \vec{x}(t+\tau) $ after $ \tau $ [s], MSD is defined as $ \langle |\Delta \vec{x}|^2\rangle_{\tau} = \langle|\vec{x}(t+\tau) - \vec{x}(t) |^2\rangle_{\tau} $.
In general, MSD is scaled as
\begin{align}
	\langle |\Delta \vec{x}|^2\rangle_{\tau} \sim \tau^{\beta},
\end{align}
where the exponent $ \beta $ reflects the trajectory of the player; in particular, $ \beta=1 $ and $ 2 $ correspond to the simple random walk and linear motion of the player, respectively.
In real data analysis, we calculated MSD for a player as the long-time average defined as follows:
\begin{align}
    \langle |\Delta \vec{x}|^2\rangle_{\tau}
    = \frac{1}{T - \tau} \sum_{t=0}^{T - \tau} |\vec{x}(t+\tau) - \vec{x}(t)|^2,
\end{align}
where $ T $ is the length of a single time series from a restart to a time-out in a game.
In Fig. \ref{msd}, we present the $ \tau $ dependence of MSD obtained from the entire time series in a game; each line in Fig. \ref{msd} is obtained by averaging MSD over all players except for the goalkeeper.
It is found that each line exhibits $ \beta=2 $ in $ \tau \lesssim 10 $ s and $ \beta=1 $ in $ \tau \gtrsim 10 $ s.
This result indicates that each player moves in a straight line for up to 10 s and then changes direction randomly.
Thus, the physics-based deterministic motion model is considered valid for a range of up to 10 seconds.

\subsection{Fujimura-Sugihara model}
We summarize the motion model proposed by Fujimura and Sugihara \cite{Fujimura2005}.
For the position $ \vec{x}(t) $ of a player at time $ t $, the Fujimura-Sugihara model is given by the following equation of motion:
\begin{align}
	m \frac{d^{2} \vec{x}(t)}{d t^{2}} &= F \vec{n} - k \frac{d \vec{x}(t)}{d t},
	\label{eq:fujimura}
\end{align}
where $ m $ is the mass of the player, $ F $ and $ \vec{n} $ are the magnitude and direction of the driving force, respectively, and $ k $ is the coefficient of viscous resistance.
In other words, the player accelerates in the direction of $ \vec{n} $ with a magnitude $ F $, and it becomes harder to accelerate in proportion to its velocity $ d\vec{x}(t)/dt $.
Given an initial position $ \vec{x}_{0}=(0, 0) $ and an initial velocity $ \vec{v}_{0} $, the solution is given as follows:
\begin{align}
	\vec{x}(t) &= \frac{1 - \exp(-\alpha t)}{\alpha} \vec{v}_{0}
	+ V_{\mathrm{max}} \left(t - \frac{1 - \exp(-\alpha t)}{\alpha}\right)\vec{n},
	\label{eq:solution}
\end{align}
where $ V_{\mathrm{max}} = F/k $ and $ \alpha = k/m $ are arbitrary constants referred to as ``kinetic parameters.''
Here, we define the first and second coefficients in Eq. \eqref{eq:solution} as
\begin{align}
	A(\alpha, t) &= \frac{1 - \exp(-\alpha t)}{\alpha}, \label{eq:A}\\[10pt]
	B(\alpha, V_{\mathrm{max}}, t) &= V_{\mathrm{max}} \left(t - \frac{1 - \exp(-\alpha t)}{\alpha}\right) = V_{\mathrm{max}} (t - A(\alpha, t)). \label{eq:B}
\end{align}
When the direction of the driving force $ \vec{n} $ changes arbitrarily, the arrival points in $ t $ [s] of the player are distributed on a circle with center $ A(\alpha, t) \vec{v}_{0} $ and radius $ B(\alpha, V_{\mathrm{max}}, t) $; this circle is referred to as ``arrival circle'' hereinafter.
Fujimura and Sugihara conducted a sprint experiment with three amateur hockey players to estimate the kinetic parameters.
By fitting the obtained speed curve with the solution of the Fujimura-Sugihara model, they empirically obtained the kinematic parameters $ V_{\mathrm{max}}=7.8 $ m/s and $ \alpha=1.3 $ 1/s as the typical values during sprinting \cite{Fujimura2005}.

\subsection{Investigation method}
\begin{figure}[htbp]
	\centering
	\includegraphics[width=6cm]{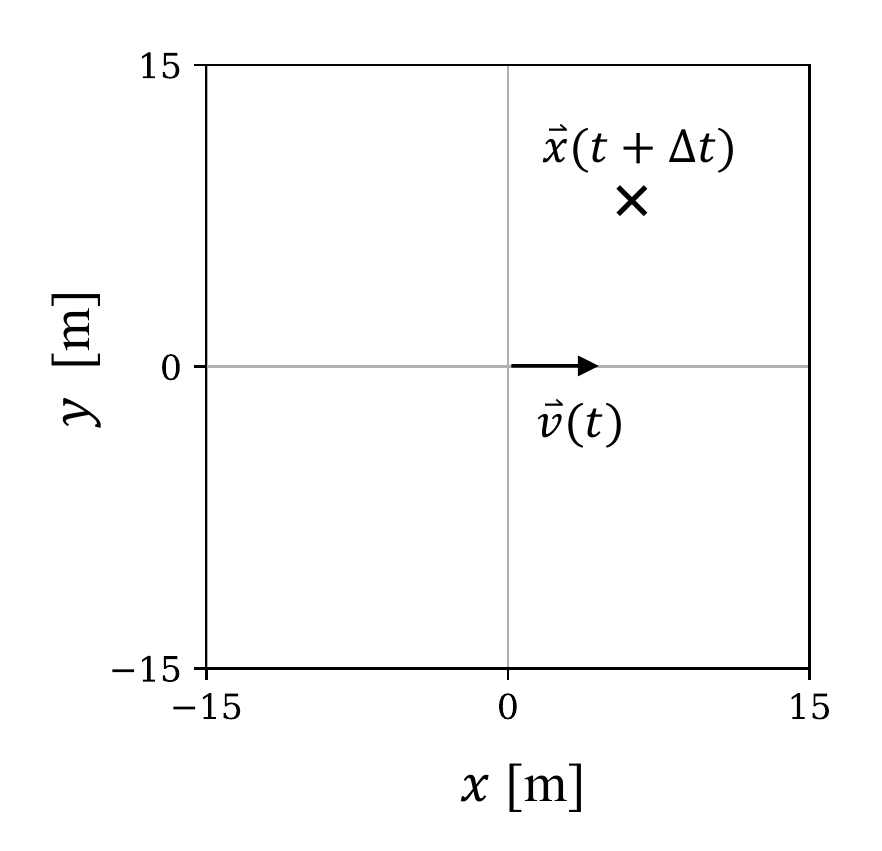}
	\caption{Coordinate system for the calculation of heat maps.}
	\label{coord}
\end{figure}

To characterize the arrival points of players in $ \Delta t $ [s] using tracking data, we adopt the method used in previous studies \cite{Fujimura2005, Brefeld2019}.
First, the velocity $ \vec{v}(t) $ of each player at time $ t $ is converted to start at the origin $ (0, 0) $ and point in the positive direction of the $ x $ axis (Fig. \ref{coord}).
We then plot the location of the same player at $ t+\Delta t $ on the same coordinates.
By repeating this plot for various players and $ t $, we obtain a heat map formed by the arrival points in $ \Delta t $ [s].
The shape of the heat map reflects the movement pattern of each player during $ \Delta t $.
For example, when players lose speed during $ \Delta t $, they arrive at a point within the heat map.
However, they arrived at a point around the boundary in the case of sprinting.
Because we investigated the validity of the Fujimura-Sugihara model under sprint conditions, we focused on the shape of the heat map's boundary and compared it with the solution \eqref{eq:solution}.
We set $ \vec{v}_{0} = (v_{0}, 0) $, where $ v_{0} > 0 $, in Eq. \eqref{eq:solution} for comparison with the heat map.
In the following analyses, we calculated heat maps using all data during the playing time of 54 games for all players except the goalkeepers.
The controlling parameters were the initial speed $ v_{0} $ [m/s] and time interval $ \Delta t $ [s].
Specifically, for a fixed $ \Delta t $, we calculated the heat maps for the initial speed $ [v_{0}, v_{0} + \Delta v_{0}) $, where $ \Delta v_{0} = 0.3 $ m/s ($ \simeq 1 $ km/h).
We set one cell side in the heat map to $ 0.2\times \Delta t $ [m], and only cells adjacent to more than $ c $ nonblank cells were used for the analyses.
To exclude isolated cells as outliers, we manually set the value of $ c $ to eight for $ \Delta t < 1 $, six for $ 1 \leq \Delta t \leq 3 $, and four for $ 3 < \Delta t $.

\section{Result}

\subsection{$ v_{0} $ dependence of heat map}
\begin{figure}[htbp]
	\centering
	\includegraphics[width=16cm]{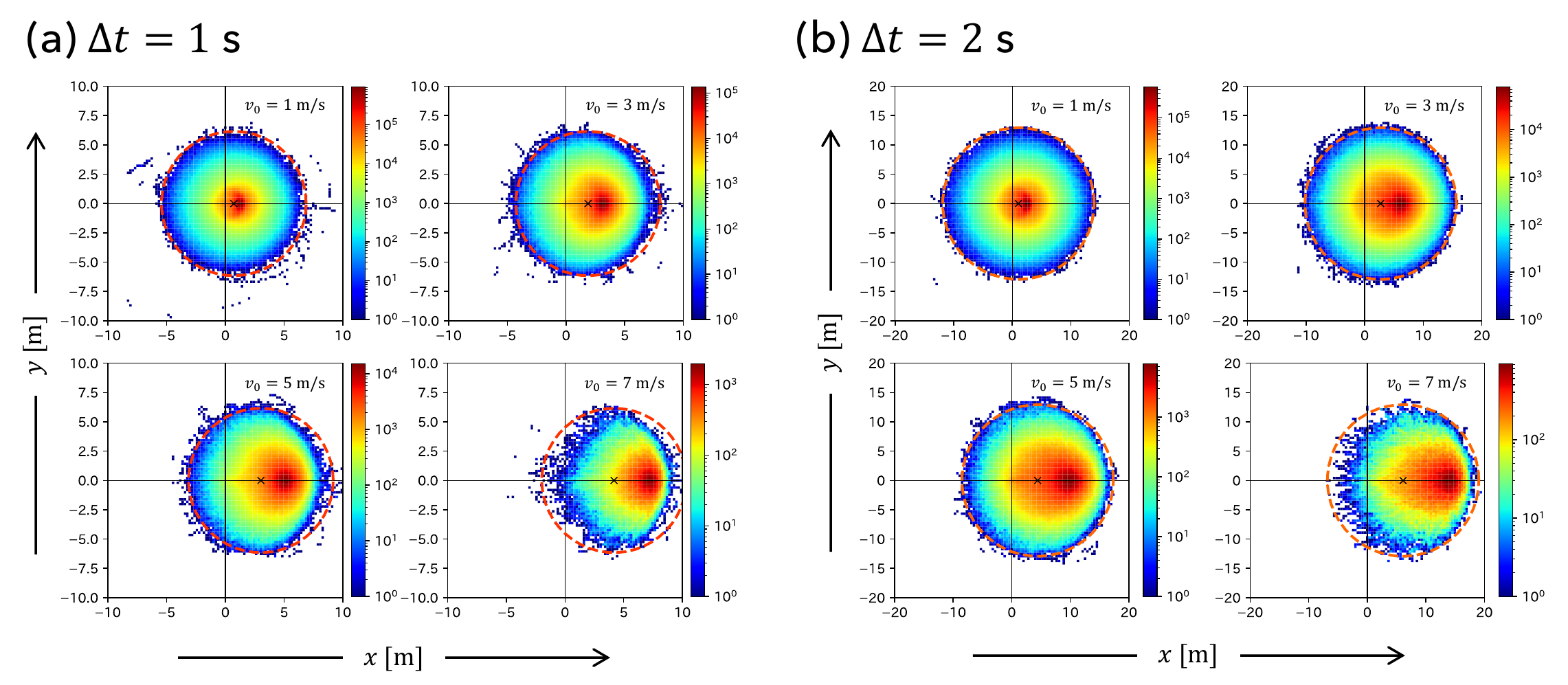} \\
	\caption{Heat maps of players' arrival points for various initial speed $ v_{0}=1, 3, 5, 7 $ m/s, where (a) $ \Delta t = 1 $ s and (b) $ \Delta t = 2 $ s. The dotted line in each panel represents the arrival circle \eqref{eq:solution} with kinetic parameters estimated from the heat maps.}
	\label{heatmap}
\end{figure}

\begin{figure}[htbp]
	\centering
	\includegraphics[width=14cm]{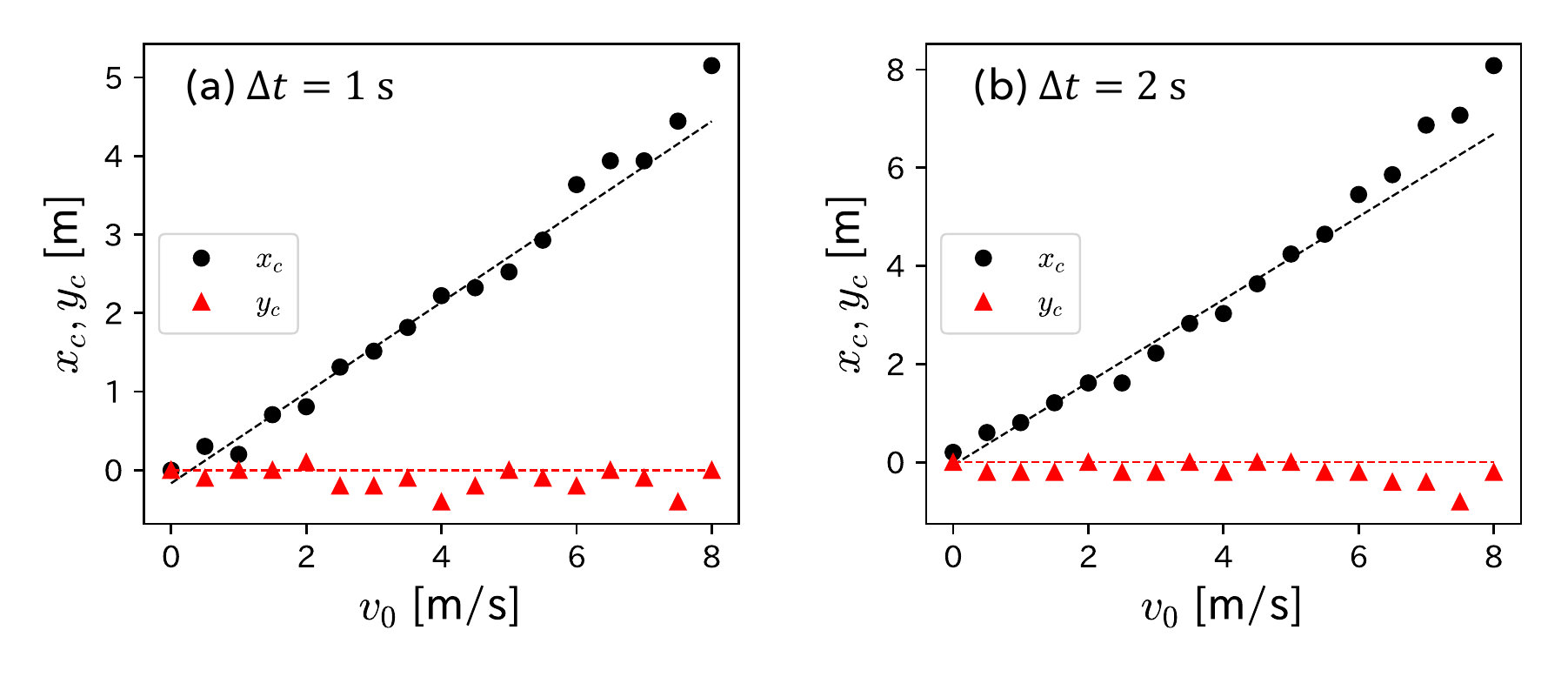}
	\caption{$ v_{0} $ dependence of the center coordinates $ x_{\mathrm{c}} $ and $ y_{\mathrm{c}} $ of the estimated circle of each heat map, where (a) $ \Delta t = 1 $ s and (b) $ \Delta t = 2 $ s. The dotted lines in each panel represent the regression line for points where $ v_{0} \leq 6 $ m/s.}
	\label{xyc-v0}
\end{figure}

\begin{figure}[htbp]
	\centering
	\includegraphics[width=14cm]{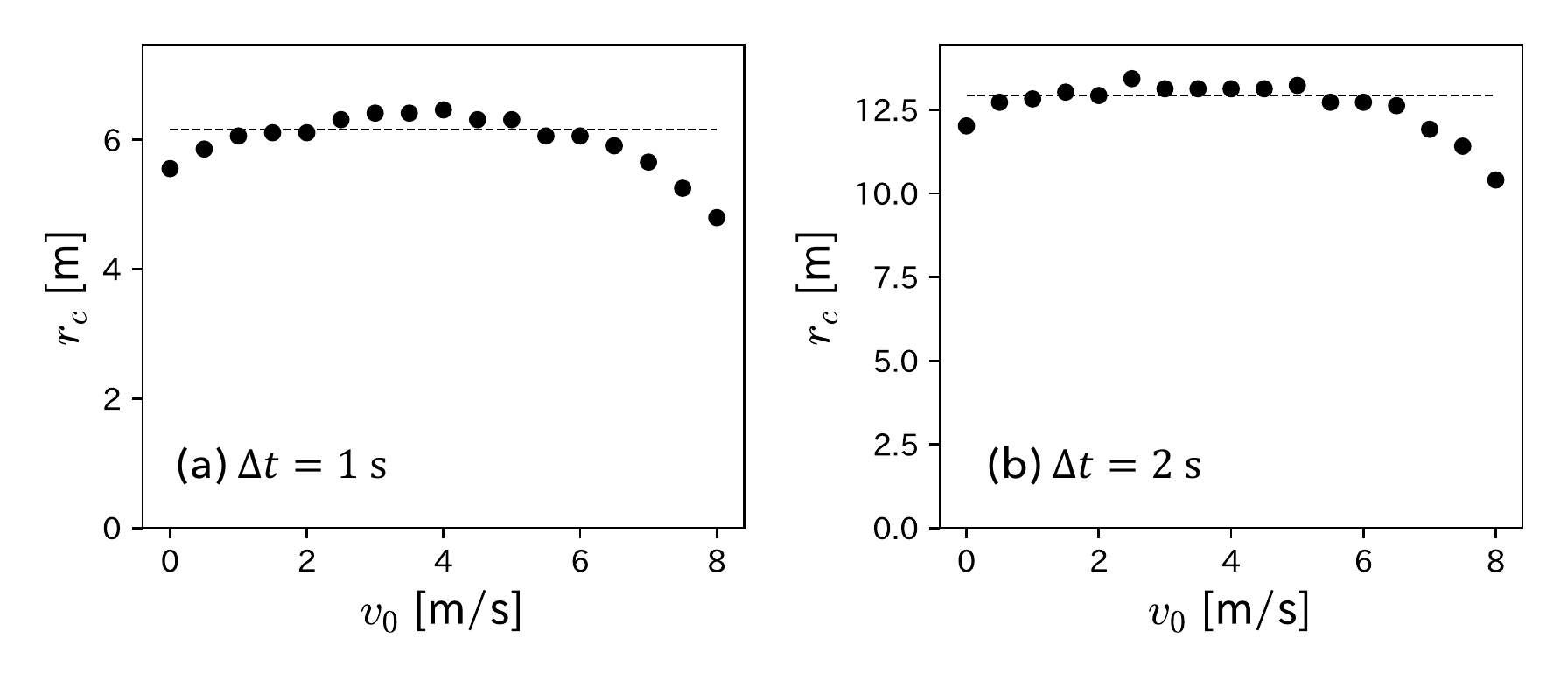}
	\caption{$ v_{0} $ dependence of the radius $ r_{\mathrm{c}} $ of the estimated circle of each heat map. The dotted line represents $ r_{\mathrm{c}} = 6.16 $ m for (a) $ \Delta t = 1 $ s, and $ r_{\mathrm{c}} = 12.94 $ m for (b) $ \Delta t = 2 $ s. These values are obtained by averaging the points where $ v_{0} \leq 6 $ m/s.}
	\label{rc-v0}
\end{figure}

Figures \ref{heatmap}(a) and \ref{heatmap}(b) present the heat maps of the players' arrival points for various initial speeds, $ v_{0}=1,\ 3,\ 5,\ 7 $ m/s, where $ \Delta t = 1 $ and $ 2 $ s.
We focused only on the shape of each heat map's boundary, although the color gradation of the heat map is proportional to the number of data points.
Remarkably, Figs. \ref{heatmap}(a) and (b) show that the boundary of the heat map is not elliptical but circular for any initial $ v_{0} $.
This result is consistent with the solution \eqref{eq:solution} of the Fujimura-Sugihara model.
Next, we approximated the boundary of each heat map as a circle and estimated its center coordinates $ (x_{\mathrm{c}}, y_{\mathrm{c}}) $ and radius $ r_{\mathrm{c}} $.
After excluding the isolated heatmap cells as outliers, we calculated the maximum and minimum values of the heat maps $ x $ and $ y $ coordinates, $ x_{\mathrm{max}} $, $ x_{\mathrm{min}}$, $ y_{\mathrm{max}} $, and $ y_{\mathrm{min}} $.
Then, $ (x_{\mathrm{c}}, y_{\mathrm{c}}) $ and $ r_{\mathrm{c}} $ were estimated as follows:
\begin{align}
	(x_{\mathrm{c}}, y_{\mathrm{c}}) &= \left(\frac{x_{\mathrm{max}}+x_{\mathrm{min}}}{2}, \frac{y_{\mathrm{max}}+y_{\mathrm{min}}}{2}\right), \\[10pt]
	r_{\mathrm{c}} &= \frac{x_{\mathrm{max}} - x_{\mathrm{min}} + y_{\mathrm{max}}-y_{\mathrm{min}}}{4}.
\end{align}
Figure \ref{xyc-v0} shows the $ v_{0} $ dependence of $ x_{\mathrm{c}} $ and $ y_{\mathrm{c}} $.
We find that $ y_{\mathrm{c}} $ is independent of $ v_{0} $; namely, $ y_{\mathrm{c}} \simeq 0 $.
However, $ x_{\mathrm{c}} $ is proportional to $ v_{0} $, particularly for $ v_{0} \lesssim 6 $ m/s.
The dotted lines in each panel of Fig. \ref{xyc-v0} represent the regression line for the points where $ v_{0} \leq 6 $ m/s.
These results are consistent with the solution \eqref{eq:solution} of the Fujimura-Sugihara model.
We also examined the $ v_{0} $ dependence of the radius $ r_{\mathrm{c}} $ of the estimated circle.
Figure \ref{rc-v0} shows that $ r_{\mathrm{c}} $ becomes almost constant, particularly for $ v_{0} \lesssim 6 $ m/s.
The dotted line in each panel of Fig. \ref{rc-v0} represents the average value of $ r_{\mathrm{c}} $ where $ v_{0} \leq 6 $ m/s.
As $ B(\alpha, V_{\mathrm{max}}, t) $ in Eq. \eqref{eq:solution} is independent of $ v_{0} $, this result is also consistent with the solution \eqref{eq:solution} of Fujimura-Sugihara model.

\subsection{Estimation of kinetic parameters using heat maps}
According to the solution \eqref{eq:solution}, the $ x $ coordinate of the arrival circle is proportional to $ v_{0} $, and the proportionality coefficient is given by Eq. \eqref{eq:A}.
Because we obtained the result that $ x_{\mathrm{c}} $ calculated from the heat map is proportional to $ v_{0} $ (refer to Fig. \ref{xyc-v0}), we can estimate the kinetic parameter $ \alpha $ using Eq. \eqref{eq:A}.
Specifically, the proportionality coefficients obtained from the regression line shown in Fig. \ref{xyc-v0}(a) and (b) are 0.58 and 0.84; then, we obtain $ \alpha=1.23 $ and $ 1.04 $ 1/s for $ \Delta t = 1 $ and $ 2 $ s, respectively.
The radius of the arrival circle is given by Eq. \eqref{eq:B}.
As shown in Fig. \ref{rc-v0}(a) and (b), the radius of the estimated circle of each heat map has become $ r_{\mathrm{c}}\simeq 6.16 $ and 12.94 m for $ \Delta t = 1 $ and $ 2 $ s.
Thus, the estimated values of $ \alpha $ and Eq. \eqref{eq:B} can yield another kinetic parameter $ V_{\mathrm{max}} $; we obtain $ V_{\mathrm{max}}=14.53 $ and $ 11.19 $ m/s, respectively.
The dotted lines in each panel of Fig. \ref{heatmap} show the arrival circle calculated by the solution \eqref{eq:solution} with the above estimated kinetic parameters $ \alpha $ and $ V_{\mathrm{max}} $.
We found that the solution \eqref{eq:solution} of the Fujimura-Sugihara model can correctly predict the boundary of the arrival region.

\subsection{$ \Delta t $ dependence of kinetic parameters}
\begin{figure}[htbp]
	\centering
	\includegraphics[width=8cm]{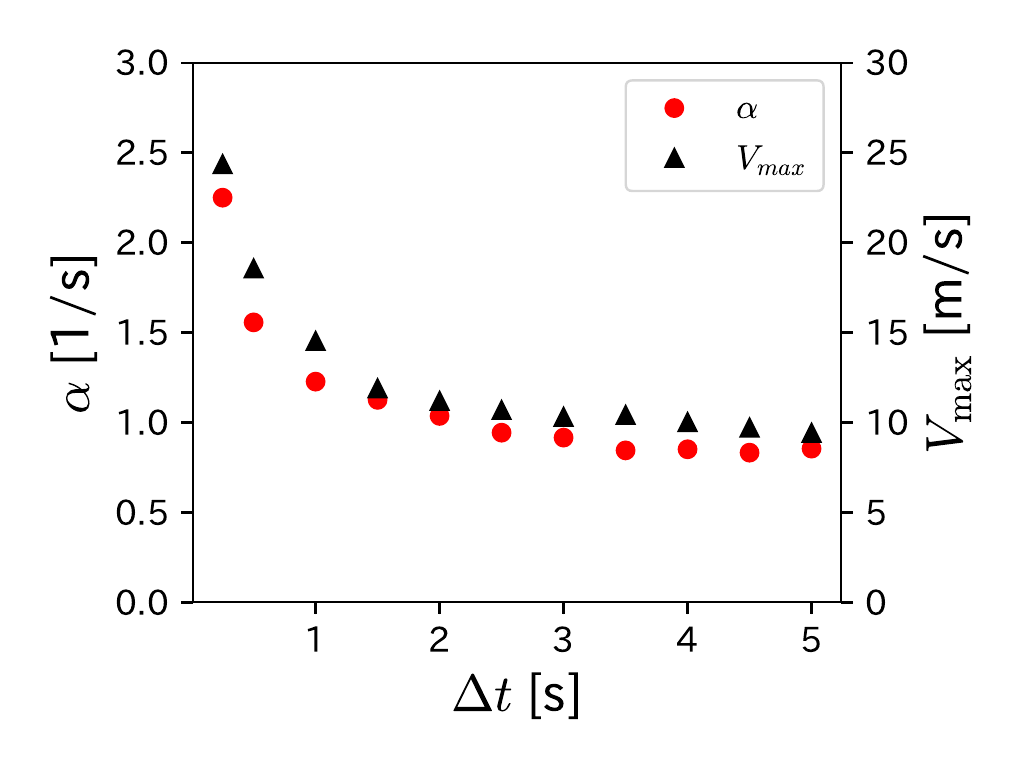}
	\caption{$ \Delta t $ dependence of kinetic parameters, $ \alpha $ and $ V_{\mathrm{max}} $.}
	\label{delta_t}
\end{figure}

We also analyze the dependence of the kinetic parameters on $ \Delta t $.
As a result of the same analyses in the previous section for different $ \Delta t $, we confirm the characteristics shown in Fig. \ref{heatmap} to Fig. \ref{rc-v0} for all $ \Delta t $.
We present the $ \Delta t $ dependence of the estimated kinetic parameters in Fig. \ref{delta_t}.
Although $ \alpha $ and $ V_{\mathrm{max}} $ are assumed to be constant in the Fujimura-Sugihara model, we found that the estimated kinetic parameters vary with $ \Delta t $, particularly in the range of $ \Delta t \lesssim 1 $ s.

\section{Discussion and Conclusion}
We investigated the validity of the Fujimura-Sugihara model based on heat maps of the players' arrival points obtained from soccer tracking data.
Our results can be summarized as follows.
First, the boundary of the heat map became a circle rather than an ellipse.
Second, $ x_{\mathrm{c}} $ was proportional to $ v_{0} $ and $ y_{\mathrm{c}} $ was independent of $ v_{0} $.
Third, $ r_{\mathrm{c}} $ was independent of $ v_{0} $.
These results are consistent with the solution \eqref{eq:solution} of the Fujimura-Sugihara model.
We also proposed a method for estimating valid kinetic parameters in the Fujimura-Sugihara model.
Meanwhile, the estimated kinetic parameters varied with $ \Delta t $, particularly in $ \Delta t \lesssim 1 $ s; this result is inconsistent with the assumption of the Fujimura-Sugihara model.
In the previous study by Fujimura and Sugihara, the kinetic parameters were estimated by a sprint experiment \cite{Fujimura2005}.
The experiment subjects were members of a college field hockey team who ran in a straight line by sprinting.
They empirically obtained the kinetic parameters $ \alpha = 1.3 $ 1/s and $ V_{\mathrm{max}}=7.8 $ m/s by fitting the solution of the Fujimura-Sugihara model to the speed curve of each subject.
In the present study, we estimated the parameters based on each soccer player's real positional data.
The obtained values were slightly different from the previous study; for example, $ \alpha = 1.04 $ 1/s and $ V_{\mathrm{max}}=11.19 $ m/s for $ \Delta t = 2 $ s.
However, our estimation method of kinetic parameters is more reasonable than the previous empirical one in the sense that the parameters were estimated directly from the soccer tracking data.
Note that as shown in Fig. \ref{heatmap} to Fig. \ref{rc-v0}, the plot in large $ v_{0} $ region is different from the others.
This discrepancy is because of the few data points of the heat maps in the large $ v_{0} $ region.
For example, a player with large $ v_{0} $ is unlikely to move in the opposite direction during $ \Delta t $.
Namely, the larger the $ v_{0} $, the fewer the data points in the region $ x < 0 $ in the heat maps.
The estimated values of $ (x_{c}, y_{c}) $ and $ r_{c} $ for large $ v_{0} $ values appear to be incorrect because they are calculated using the maximum and minimum values of the heat maps $ x $ and $ y $ coordinates.
In the future, the use of large-scale tracking data could eliminate such problems.
We comment on the results of the $ \Delta t $ dependence of the kinetic parameters shown in Fig \ref{delta_t}.
This result requires careful consideration from two perspectives.
First, the $ x $ and $ y $ coordinates of our tracking data contain an error of $ \pm 1 $ m by assessing the accuracy of the TRACAB system.
This error cannot be ignored in the analysis of the heat maps, especially when $ \Delta t \lesssim 1 $ s.
Therefore, we must check the validity of the rapid increase in $ \alpha $ and $ V_{\mathrm{max}} $ in Fig. \ref{delta_t} using more accurate data.
The second possibility is that the results in Fig. \ref{delta_t} indicate the limitation of the Fujimura-Sugihara model.
The Fujimura-Sugihara model must be extended to reproduce the behavior shown in Fig. \ref{delta_t}.
One possible extension is a model in which the magnitudes of the driving force $ F $ and viscous resistance $ k $ in Eq. \eqref{eq:fujimura} are time-dependent:
\begin{align}
	m \frac{d^{2} \vec{x}(t)}{d t^{2}} &= F(t) \vec{n} - k(t) \frac{d \vec{x}(t)}{d t}.
	\label{eq:fujimura2}
\end{align}
This equation is known as the variable coefficient second-order linear ODE.
As the time dependence of kinetic parameters is conceivable in soccer, the analysis of this extended motion model can be a challenging and significant future topic in soccer game analysis.
Previously, Brefeld et al. demonstrated that the arrival region of soccer players becomes elliptical \cite{Brefeld2019}.
Fern\'{a}ndez et al. modeled the player influence area as elliptical when the initial speed is large \cite{Fernandez2018}.
However, Anzer et al. pointed out that the smaller the speed interval $ \Delta v_{0} $, the closer the shape of the arrival region is to be circular rather than elliptical \cite{Anzer2022}.
This study supports the results of Anzer et al. with a more detailed and comprehensive analysis.
Furthermore, we have recently shown that soccer players' sprints satisfy the characteristics of the Fujimura-Sugihara model, that is, $ v_{0} $ dependence of the arrival circle for fixed $ \Delta t $.
Our results suggest that a relatively simple model can describe soccer players' sprints, although a slight discrepancy exists between actual observations and model predictions.
Furthermore, the motion model under the sprint condition was utilized for pass prediction \cite{Fujimura2005, Spearman2017, Alguacil2020} and for modeling the dominance and influence of soccer players at each location in the field \cite{Narizuka2021}.
Our observations are fundamental characteristics that the motion model should satisfy and provide direction for modeling soccer players' motions.
In conclusion, the Fujimura-Sugihara model effectively predicts the arrival point of soccer players by sprinting, except when the time interval $ \Delta t $ is small.
The boundary of the player's arrival region is shown to be circular rather than elliptical; the initial speed dependence of the arrival region satisfies the model.
In the case of sprinting, kinetic parameters in the model can be estimated directly from the soccer tracking data.

\section*{Acknowledgements}
The authors are grateful to DataStadium Inc., Japan, for providing the player tracking data for this study.
This work was partially supported by the Data-Centric Science Research Commons Project of the Research Organization of Information and Systems, Japan, a Grant-in-Aid for Young Scientists (18K18013) from the Japan Society for the Promotion of Science (JSPS), and Hayao Nakayama Foundation for Science and Technology and Culture (22-KI-09).
%

\bibliography{./reference}

\end{document}